\documentclass[12pt]{article}
\usepackage[dvips]{graphicx}
\begin{document}

\title{
Effects of stochastic nucleation in the first
order phase transition}
\author{Victor Kurasov}
\date{Victor.Kurasov@pobox.spbu.ru}

\maketitle

In the system during the first order phase
transition  new  supercritical
formations of a new phase appear
with some fixed probability, but appear in stochastic
manner.
In a system with macroscopic sizes  due to a giant
value of the Avogadro number there appears rather
big number of droplets.
It allows to use the averaged characteristics to
construct  kinetics of a nucleation process.
Kinetics based on averaged  characteristics
is described in \cite{PhysRevE94}.
In this paper the time evolution is constructed and
it is possible to extract elementary intervals
where thermodynamic  parameters and the nucleation rate
have small variations.
The total number of droplets is so big that
at every elementary interval there appears a great
number of droplets   $\Delta N$.
On the base of traditional thermodynamics
one can state that the relative fluctuation
 $\delta \Delta N/ \Delta N$
of droplets formed at elementary interval is
small and has an order of
$(\Delta N)^{1/2}$.
This remark completely solves a problem of
justification of nucleation description based on
averaged characteristics.
But recently there appears a set of papers \cite{Vest},
\cite{Koll}
where a stochastic effects (the effects of
fluctuations of droplets formation) were
described and investigated.
Two approaches were
formulated there. Authors didn't hesitate
that these approaches \cite{Koll} , \cite{Vest}
gave different results.
So, one has to analyze approaches \cite{Vest}, \cite{Koll}
and decide what is the true result.

An idea formulated in \cite{Koll},  \cite{Vest},
allows to establish corrections to the total
number of droplets $N$
appeared in the system.
It was supposed that these corrections  are
functions of
$N$.
To demonstrate the error of this approach it is
sufficient to take two identical systems then to
calculate them separately and to add results or
to calculate correction directly for the total
system. The results will  differ.

One has to determine a real volume to which one
has to refer the number of droplets.
It is simple to do with the help of results from
\cite{PhysicaA}.
In that paper  kinetics of nucleation for
diffusion regime of droplets growth was
constructed. It was shown that a solitary droplet
perturbes vapor up to distances of an order
$\sqrt{4Dt}$,
where $D$
 is diffusion coefficient,
 $t$
 is a time from  droplet formation. One can take as
    $t$
 a time
  $t_1$
  of nucleation period duration. A nucleation
  period is a period of intensive formation of
  droplets.

It allows to give a new definition of the volume
 $V_1$
where the number  $N_1$ of droplets is formed.
Namely this value has to be regarded as a volume
of a system.
This volume is $4\pi (4D t_1)^{3/2}/3$.
If the sizes of the system are smaller than this value
one has to take the volume of the system as this
value.
But such a small system can be hardly regarded as
a macroscopic one. At least one has to analyze
carefully the boundary conditions.

Naturally, the droplets appeared at different
times perturbe initial phase at different
distances. So, one can regard above formulas only
as estimates. Some more rigorous equations can be
found in
\cite{book2}.

The number of droplets  $N_1$
isn't too big as
$N$ is.
So, an analysis of stochastic effects has a real
sense. It is interesting now to get all
correction terms which are ascending with the
number of droplets (but not  only a leading
term).
To solve this task one has to modify approaches from
\cite{Koll}, \cite{Vest} where only a vague conclusion
about the order of the leading term was made.

Complexity of this problem lies in conclusion
that one can not directly use equations based on
the theory with averaged characteristics. In
\cite{Koll},
\cite{Vest}
some properties of solution on the base of
averaged characteristics were starting points for
constructions.
This supposition was adopted without any
justifications.

We shall consider the situation of decay. The new
dimensionless
parameter - the number of droplets destroys the
universality observed in
\cite{PhysRevE94}
for the theory based on averaged characteristics.
Moreover, it is difficult even to formulate the
system of equations. It radically complicates the
problem.

The property of effective monodispersity
formulated in
\cite{Mono}
was used in  \cite{Vest}
without any justifications. Generally speaking
this property can not be directly used to
calculate
stochastic corrections. This leads to an
error made in
 \cite{Vest}.
 Now we shall formulate more correct approach.

Both approaches from \cite{Koll} and
\cite{Vest} used the following property
"The droplets formed at the beginning of the
nucleation period are the main consumers of
vapor".
This property is valid, but it is substituted in
 \cite{Vest}, \cite{Koll}
by the following statement:
"The main source of stochastic effects are the
free fluctuations of
droplets formed at the beginning of the
nucleation period. they govern the fluctuations
of all other droplets". But the last statement isn't
valid.
So, one has to use some new methods which are
presented below.

The use of monodisperse
approximation will lead to some errors. But due to
universality of solution \cite{PhysRevE94}
these errors  can not be be too big.  Qualitatively
everything is suitable, but precision
will  be very low.

The same conclusion will be valid
for calculations based on some model
behavior of supersaturation (justification is
valid for a vapor consumption, but not for
stochastic effects).
Here the final result will be more precise but it
comes from rather spontaneous artificial choice
of some parameter which equals in
\cite{Koll}
to $1/2$.
In \cite{Koll}
it is supposed that until some moment
(it is chosen in  \cite{Koll}
as a half of nucleation period) the droplets
are formed under ideal conditions and namely
these droplets determine a vapor consumption.
This approach taken from
\cite{book1},
was used in  \cite{Koll}
in slightly  another sense.
It is supposed that droplets formed during a
 first half of nucleation period
 are the main source of stochastic effects.
The last statement was not justified in  \cite{Koll}
and it is rather approximate. The relative
correctness  of a result was attained due to
specific compensation of different errors of
approximations used in
\cite{Koll}.

All arguments listed above lead to necessity of
reconsideration which will be made in this
publication.
A plan will be the following
\begin{itemize}
\item
On the basis of algebraic approach we shall see
that stochastic effects are small
\item
A smallness of stochastic effects allows to seek
the solution on the base of the theory with
averaged characteristics. But we have to take
stochastic effects from all droplets formed
during the nucleation period.
\item
The possibility to take into account the role of
all droplets can be ensured by the property of
similarity of nucleation conditions during  the
nucleation period. This property can be
considered in two senses - 1) in the local
differential sense  and 2) in the integral sense
in frames of the first iteration. The local
property will be used in justification of the
smallness of stochastic effects and the integral
property will be used to calculate the
stochastic corrections.
\end{itemize}

All analytical results will be checked by
computer simulation and a coincidence  will be shown

All mentioned constructions will be valid for an
arbitrary first order phase transition.

The law of droplets growth will be a free
molecular one, the linear size grows with
velocity independent from its value.
Consideration of other regimes can be attained in
frames of the current approach by some trivial
substitutions, but one has to take into account
that the new regime requires new approaches to
construct nucleation kinetics as it is shown in
 \cite{PhysicaA}.
So, we can not agree with he statement in \cite{Vest}
that one of results is an account of stochastic
effects in a diffusion regime of droplets growth.
This effect has to be taken into account by
application of methods presented in
\cite{PhysRevE2001}.

Dynamic conditions can be easily considered by
direct generalization of methods presented here
and we needn't to present it in details.

\section{Estimates for stochastic effects}

We consider the theory based
on averaged characteristics.
It is supposed to be known \cite{PhysRevE94},
that the supersaturation $\zeta$
behavior can be determined after
certain renormalizations by the following equation
$$
\psi (z) = \int_0^t dx (z-x)^3
\exp(-\psi)
$$
A good approximation for solution and for
a distribution  $\exp(\psi)$
over linear sizes is
$f_1 =
\exp(-z^4/4)$.
The form of $f_1$
is given by fig.1
\begin{figure}[hgh]

\includegraphics[angle=270,totalheight=10cm]{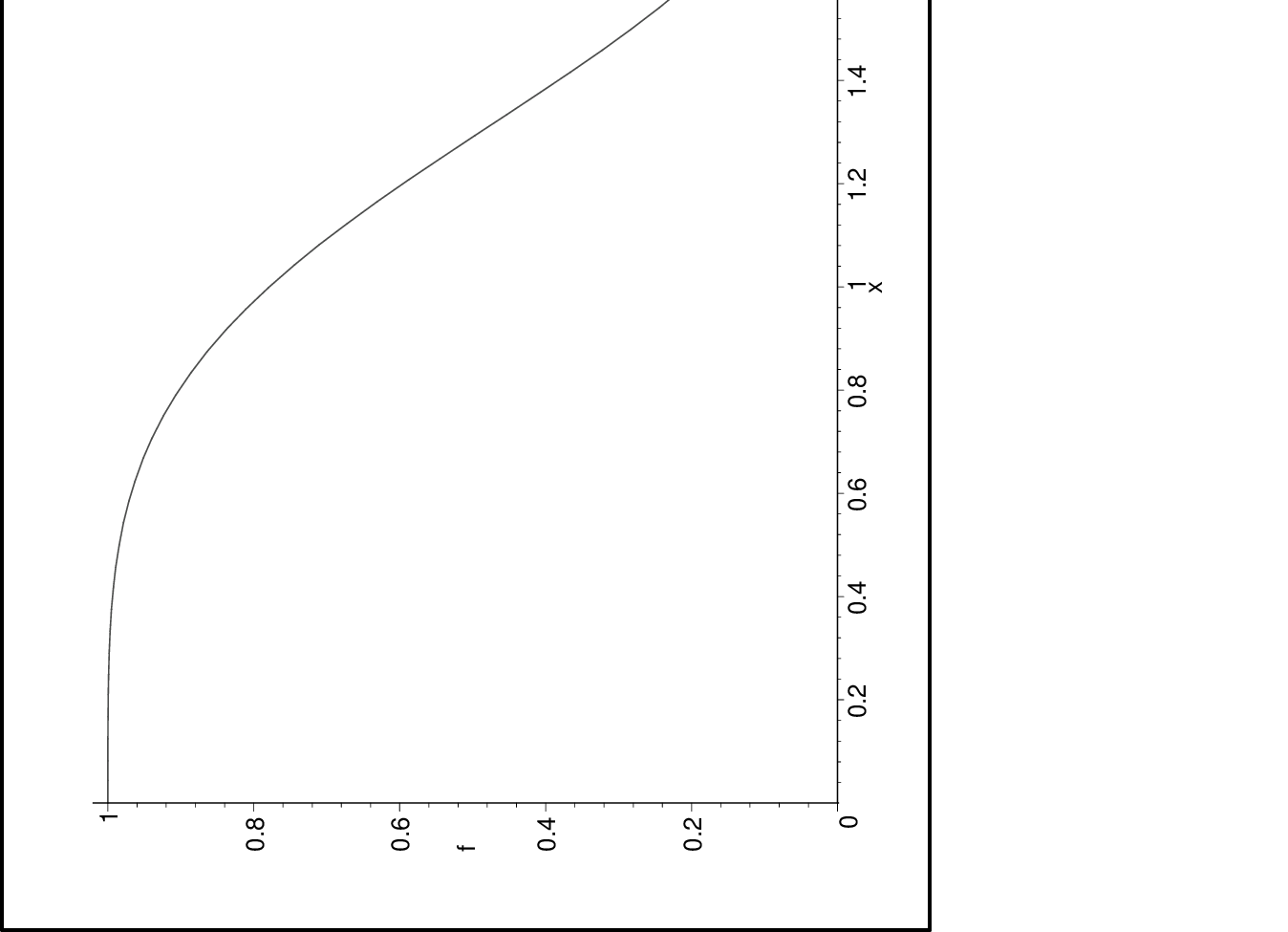}

\begin{caption}
{
A form of approximation for size spectrum}
\end{caption}

\end{figure}
It is seen that  at $z_0=1.25$
the nucleation period stops.

This approximation has rather high precision
\cite{Mono}.
 It is based on the following law of substance
accumulation
$$
G = z^4/4 = \int_0^z (z-x)^3 \equiv \int_0^z g, \
\ g = (z-x)^3
$$
For any moment $t$
or $z$
a function $g$
has one and the same form.
This is a similarity of nucleation
conditions. We see that every time the droplets
formed at the last third of a period from
beginning of nucleation until a current moment
will accumulate a negligible quantity of
substance.
The relative quantity of the substance there has
an order of
($ \sim 1/27$)
and is so small that even if there will be
fluctuations the quantity will be small.

From the form  of $f_1$
it is seen that until
$z_= \equiv 0.7 z_0$
all droplets will deplete vapor rather weak. It
will be important for future analysis.

The mentioned property of $g$
allows to use a monodisperse approximation
\cite{Mono}
not only at the end of nucleation but in every
moment of the nucleation period
\cite{Mono}.
Let $t(G)$
be the moment when there are $G$ molecules in droplets
(in appropriate units).
An application of the monodisperse approximation
 \cite{Mono}
leads to
$$
\frac{1}{4} N(z) z^3 = G
$$

Now it will be possible to repeat all
constructions   \cite{Vest}
with $G$ instead of $1$
(in renormalized units, before renormalization
it would be $1/\Gamma$  (see \cite{PhysRevE94}).
The sense of these transformations is rather evident.
Let $t(z)$ be the current moment of time
  ($z$ is the coordinate of ta spectrum front,
actually $t$ is proportional  $z$).
We suppose that before $az$ ($a$
 is some parameter)
droplets are formed without mutual influence and
one can write Poisson's distribution.
Then a natural restriction on
$a$
appeared, namely $a< 0.7*z_0/z$.
Then we suppose that the influence of other droplets
on its own formation is negligible
(this follows from
$2 * 0.7 = 1.4 > 1.25$
and from notation about the last third of
nucleation period).
Then it is possible to write Poisson's
distribution for the second group of droplets,
but with parameters depended on stochastic
values - characteristics of the droplets
distribution from the first group. Rigorously
speaking one has to use
the first four moments of the droplets distribution
in accordance with \cite{PhysRevE94}, but for
simplicity we shall use here only the zero
momentum. As a compensation for this simplicity
we has to use here
$a=0.25$.

Then one has to come from
Poisson's distributions to Gauss distributions
and integrate them with account of connection
between stochastic parameters of embryos
formation from the first group and parameters of
distribution from the second group.
Unfortunately, in
\cite{Vest}
the final result was an expression for
distribution, but not for the number of droplets
which can be observed in experiment. Beside this
the expression for the distribution was
calculated only in a leading term which is
certainly a Gauss distribution. Corrections
haven't been established.

Contrary to \cite{Vest}
we shall take into account all correction terms
which comes from transition from Poisson's
distributions to Gauss distributions and corrections for
nonlinear connection between group distributions.
We shall take all terms which are growing when
the total number of droplets grows.
We shall get the following result for droplets
distribution
$$
P = (\frac{9 a}{2 \pi N (3a+1)})^{1/2}
\exp(-\frac{9a}{3a+1} \frac{D_s^2}{2})
(1+ y)
$$
where
$$
D_s = \frac{\hat{N}-N}{\sqrt{N}}
$$
$\hat{N}$ -
 some stochastic  value of the total number of
 droplets,
$N$
- the mean value of droplets and $y$ is the
 correction for spectrum.

At $à= 1/4$
we get
$$
y=
\frac{1}{74088} D_s (8087 D_s^2 - 10269) s
+ (-\frac{4}{9} + \frac{305}{1176} D_s^2
$$
$$
-\frac{8503}{1176}D_s^2 -
\frac{85903}{12446784} D_s^4 +
\frac{65399569}{10978063488} D_s^6 ) s^2
$$
where
$$
s \equiv \sqrt{N}
$$
is a small parameter of decomposition.
To
get all ascending corrections we must fulfill
decomposition until
$s^2$.
At arbitrary  $a$
we get the following expression
$$
1+ w_1 s+ w_2 s^2
$$
Here
$$
w_1 =
-\frac{1}{6} (( 486 l^2 + 486 D_s^2 l^{12} -
972 l^{11} D_s^2 + 324 l^4 - 648 D_s^2 l^{10}
+ 756 l^9
$$
$$
+ 810 l^9 D_s^2 - 459 l^8 + 297 D_s^2 l^8 -
135 l^7  D_s^2 -
$$
$$
387 l^7 + 90 l^6 D_s^2 - 153 l^6
+ 27 l^5 D_s^2 - 213 l^5 - 66 l^4 D_s^2 -
3 l^4  + 27 l^3 - 3 l^3 D_s^2 + 16 l^2 D_s^2 - 3
l^2 +
$$
$$
9 l + l D_s^2 + D_s^2) D_s ) /
((1+ 3 l^2)^3 (l+1) (-1+l^2))
$$
where
$$
l=\sqrt{a}
$$
and
$$
w_2 = w_{02} / ((l+1)^2 (-1+l^2) (1+3 l^2)^6 l^2
)
$$
$$
w_{02} =  \sum_{i=0}^3 q_{2i} D_s^{2i}
$$
$$
q_{0} = - \frac{1}{12} - \frac{1}{6} l +
\frac{243}{4} l^{16} - \frac{135}{2} l^7 - 90 l^8
- \frac{87}{2} l^6 + \frac{243}{2} l^{14} +
\frac{81}{2} l^{12}
$$
$$
- \frac{153}{2} l^{10} -
\frac{67}{6} l^4
-\frac{225}{2} l^9 - \frac{3}{2} l^2 -
\frac{81}{2} l^{11} + \frac{243}{2} l^{13} -
\frac{243}{2} l^{15} - \frac{17}{6} l^3
$$
$$
q_2 =
\frac{5265}{4} l^{12} + \frac{51}{2} l^6 + 117
l^{10} + \frac{333}{2} l^{11} + \frac{31}{8} l^4
$$
$$
+ \frac{17}{2} l^5 + \frac{5049}{2} l^{14} -
\frac{15309}{2} l^{19} + 2079 l^{13}
$$
$$
+ 31 l^8 -
8748 l^{18} + 42 l^9  + \frac{24057}{2} l^{21} +
\frac{103}{2} l^7 + \frac{16767}{8} l^{20} + 6561
l^{22}
$$
$$
+ \frac{4617}{2} l^{15} - 3159 l^{16} -
8262 l^{17}
$$
$$
q_4 =
- \frac{4413}{2} l^{13} - \frac{477}{2} l^{15} +
7047 l^{21} - \frac{1161}{4} l^{18}
$$
$$
+\frac{361}{6} l^8 +\frac{105}{2} l^{11} -
\frac{104247}{4} l^{22} - \frac{3}{4} l^4 + 17091
l^{20}
$$
$$
- 5103 l^{23} + \frac{43}{6} l^5 -
\frac{31185}{4} l^{19} + 7965 l^{17} -
\frac{3213}{4} l^{12} - \frac{475}{3} l^{10} +
510 l^{14} - \frac{49}{6} l^7 + \frac{23}{4} l^6
$$
$$
+ \frac{1}{12} l^3 + {44469}{4} l^{24} -
\frac{1}{12} l^2 - \frac{6903}{4} l^{16} +
\frac{74}{3} l^9
$$
$$
q_6 =
- 13122 l^{25} - \frac{40419}{8} l^{18} +
\frac{17253}{2} l^{20} - \frac{24057}{2} l^{22} +
4374 l^{24} + \frac{6561}{2} l^{26} -
$$
$$
\frac{48843}{2} l^{21} - \frac{623}{24} l^8 +
\frac{59}{36} l^6 - \frac{3609}{4} l^{14}
$$
$$
-\frac{57}{8} l^{12} + \frac{411}{4} l^{10} +
\frac{11}{24} l^4 + \frac{1}{72} l^2 -
\frac{29}{12} l^7
-\frac{345}{4} l^{11} + \frac{1}{36} l^3  +
$$
$$
\frac{13}{36} l^5  + \frac{65}{4} l^9 +
\frac{2421}{4} l^{13} - \frac{7857}{4} l^{15} +
$$
$$
\frac{8667}{4} l^{17} + 28431 l^{23} + 7047
l^{19} + \frac{24705}{8} l^{16}
$$
Having integrated this expression we get
corrections to  droplets number.
The term at $s$
gives zero after integration and the first correction
has an order of $\sim s^2$
and doesn't depend on the total number of droplets.
A coefficient at $s^2$
has at $a=1/4$
a value $$ d_0 =311/3024 $$
At arbitrary $a$
a coefficient at $s^2$
in correction for the total number of droplets
will be
$$
d_a =
\frac{1}{72} [ 108 a^6 + 540 a^{11/2}
- 72 a^5 - 930 a^{9/2} - 336 a^4 +
$$
$$
713 a^{7/2} + 158 a^3 +4 a^2 - 203 a^{5/2}
- 6 a + 39 a^{3/2} - 3 a^{1/2} ] /
[ a^{3/2} (1-a)^2 (1+3a)]
$$

It will be interesting to compare results with
and without corrections from transition from
Poisson's distribution to Gauss distribution.
At the leading term there will be no change. At
correction terms we
have
$$
y =
\frac{17}{74088} D_s (-2331+289D_s^2)s +
(\frac{17}{196} D_s^2 +
$$
$$
\frac{732037}{12446784} D_s^4 +
\frac{24137569}{10978063488} D_s^6 -
\frac{13}{36}) s^2
$$
$$
d_0 =
-37/126
$$
$$
d_a =
\frac{1}{72}
( 648 a^{11/2}  - 216 a^5 - 1062 a^{9/2}
+108 a^4 + 753 a^{7/2} -
$$
$$
30 a^3 -
195 a^{5/2} - 12 a^2 +19 a^{3/2} + 6 a
- 7 a^{1/2} ) /
( a^{3/2} (1+3a) (1-a)^2 )
$$

It is seen that these corrections are small. At
arbitrary $a$ except too small and $1$ (these
values are unreal) we get values shown at fig.2

\begin{figure}[hgh]

\includegraphics[angle=270,totalheight=10cm]{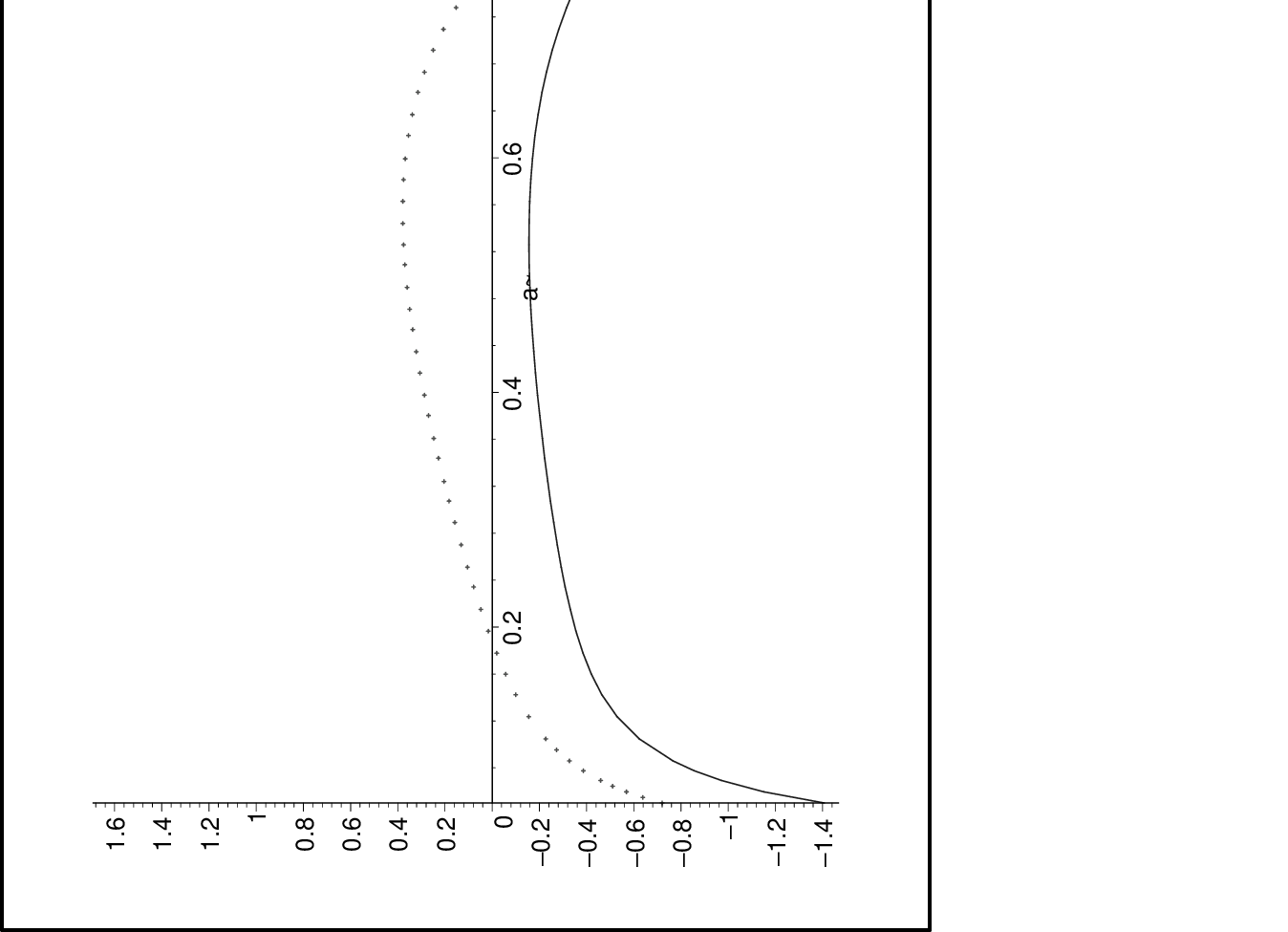}

\begin{caption}
{
Dependence of corrections on $a$}
\end{caption}
\end{figure}

Points show corrections with transition
from Poisson's distribution to Gauss
distribution,  a line  shows corrections without transition
from Poisson's distribution to Gauss
distribution. Both corrections have one and the
same order and they are small.

One has to note that we have not take into
account corrections from transition from
summation to integration.
It is made due to reasons
formulated below.
Really, we have at transition from summation in
formula
$$
P(N) = \sum_{\hat{N_1}} P_1 (\hat{N_1}, N_1) P_2
(N-\hat{N_1}, N_2(N_1))
$$
to integration
$$
P(N) = \int d \hat{N_1} P_1 (\hat{N_1}, N_1) P_2
(N-\hat{N_1}, N_2(N_1))
$$
to use Euler's decomposition.
But discrete character in nucleation isn't so
trivial. The process of vapor consumption
can not begin
without the first droplet.
The system will wait for droplet as long as it
will be necessary.
It shows that discrete effects are complicate and
require a separate publication.

To use Poisson's distribution for the first group
one has to make the following notation. Really
nucleation conditions for the first group don't
differ from the whole group. So, for distribution
for the first group one has to take distribution
$P_1$ with reduced halfwidth.
But one can not attribute a halfwidth to
Poisson's distribution. That's why we considered
effects with and without corrections from
transition from Poisson's to Gauss distribution.
So, we can use Gauss distributions as initial
ones.
For Gauss distribution one can easily
reconsider the halfwidth. Then for
$P_1$
one can take
$$
P_1 \sim \exp((\hat{N_1} - N_1)^2 / (2 b  N_1))
$$
where $b$
is a renormalization coefficient.
Distribution $P_2$ remains previous
$$
P_2 \sim
\exp((\hat{N_2} - N_2 )^2 / (2   N_2))
$$
where $N_2$
is given by
$$
N_2 = (1-\frac{1}{3} s + \frac{2}{9} s^2
- \frac{14}{81} s^3 + \frac{35}{243} s^4 -
\frac{91}{729} s^5 - a)N
$$
where
$N$ is a mean total number of droplets,
$$
s = \frac{\hat{N_1} - a N}{a N }
$$
is a small parameter of an order
$N^{-1/2}$

After integration one comes to
$$
P \sim \exp( -
\frac{9 a b^2}
{2(9 a  + b - 6 b a + 9 b a^2  -
9 a^2)}d^2)
$$
where
$$
d= (\hat{N} - N)/\sqrt{N}
$$

The halfwidth of the distribution $P_1$
  must be equal to the halfwidth of $P$,
  which leads to
 $$
 b= 9 \frac{a(1-a)}{-9 a^2 +15 a - 1}
 $$

Now we shall consider  effects of
renormalization. The ratio of corrections with
renormalization and without renormalizations is
given by
$$
r = \frac{1}{9}
\frac{\sqrt{-9 a^2 + 15 a -1}
\sqrt{3 a +1}}{a \sqrt{1-a}}
$$
and it is shown at fig. 3

\begin{figure}[hgh]

\includegraphics[angle=270,totalheight=10cm]{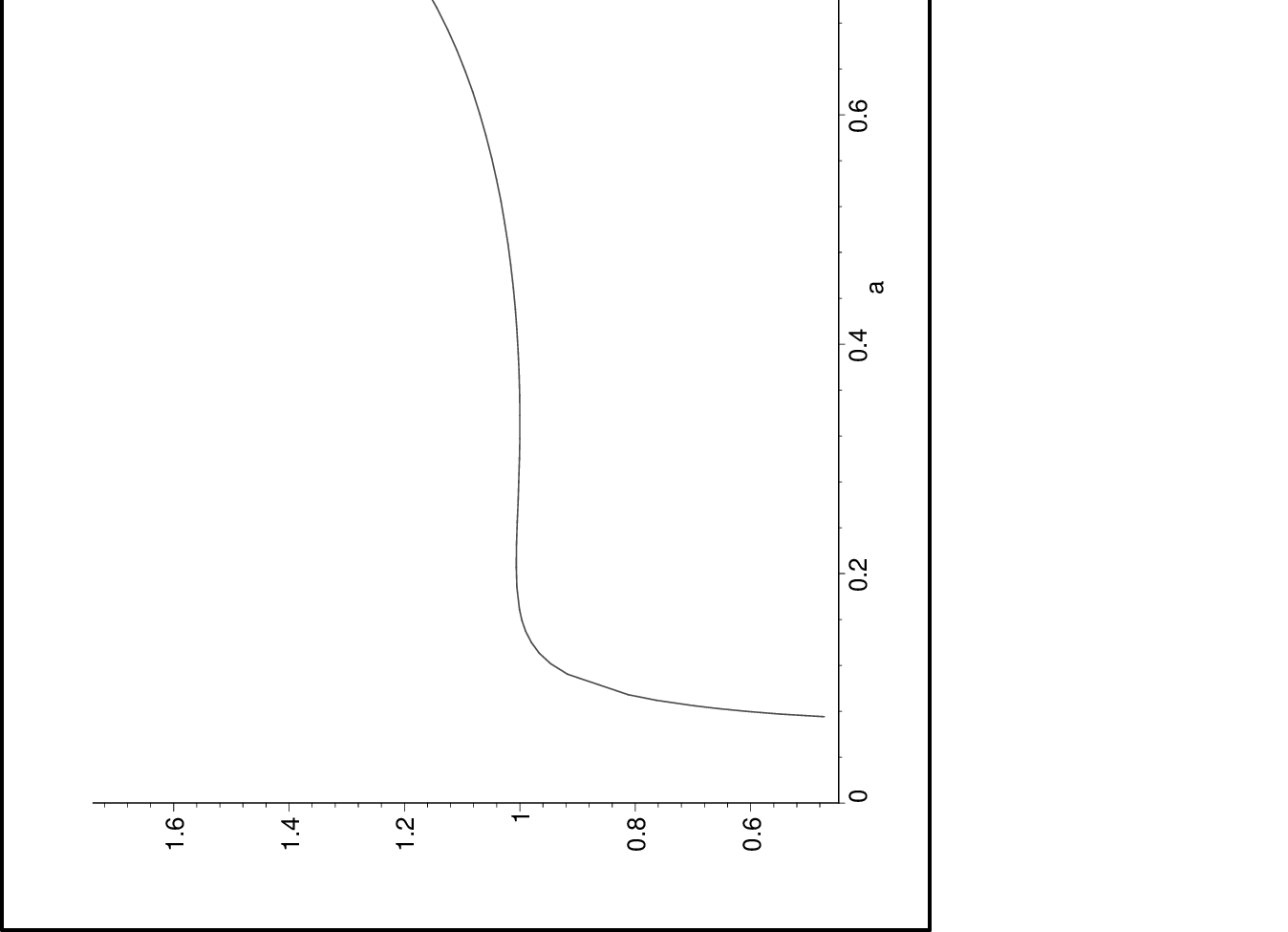}

\begin{caption}
{
Ratio of halfwidths
}
\end{caption}

\end{figure}

For all reasonable values of $a$
it is approximately  $1$. At $a=1/4$
 we get
$1.0041$
So, here the effect of similarity of nucleation
conditions doesn't lead to
remarkable effects.
It is only due to monodisperse approximation.
Later this effect will be essential.

Instead of taking into account all moments of
distribution we can directly calculate the effects
on the base of explicit form of spectrum in
frames of iteration procedure.

The result of this section is the following: we
have proved that stochastic effects are small.
Beside this we have demonstrated how to use the
similarity of nucleation conditions. Here it is
useless, but later this method will lead to some
essential numerical results.

\section{
Stochastic effects in the iteration procedure}

There is  only one parameter which is essentially
deviated by stochastic effects - it is dispersion
of distribution or a halfwidth of  distribution.
Now we shall calculate this value.

The most advanced approach was suggested in
\cite{Koll}.
But even this approach has many disadvantages.

We shall characterize a droplet by linear size $\rho$
which is the cubic root
of its molecules number. Its velocity of growth
does not depend on $\rho$.

Decomposition of a whole interval of nucleation
into elementary intervals is connected with some
difficulties.
An elementary length
$\tilde{\Delta}$
must satisfy according to \cite{Koll} to
two requirements:
1. A number of droplets formed during elementary
length must be big.
2. An amplitude of a spectrum has to be
approximately constant during an elementary
interval.

It is clear that the second requirement
can not be satisfied. Stochastic deviations of an
amplitude leads to
violation of the second requirement.

We shall apply the second requirement not to stochastic
amplitude as it was stated in \cite{Koll},
but to averaged amplitude. Then the second
requirement is : An averaged
amplitude of a spectrum has to be
approximately constant during an elementary
interval.

Stochastic amplitudes  $f_i$ are
introduced in \cite{Koll} as
$$
f_i = \frac{N_i}{\tilde{\Delta}}
$$
where
$N_i$
is the number of droplets
formed during
$\tilde{\Delta}$. It isn't stochastic value but
partially averaged one.
An expression for the number of molecules
in droplets formed during interval number $i$
at a moment  $t_k $ (it means that now we are at
interval number $k$)
will be the following one
$$
\int_{x_{k-i}}^{x_{k-i+1}} f_i \rho^3 d \rho =
\frac{1}{4} f_i (x_{k-i+1}^4 - x_{k-i}^4)
$$
The difference between forth powers corresponds
to a constant amplitude of spectrum. It is wrong
and then eq. (12) in \cite{Koll} and all further
equations  are not correct.

But the is no necessity to use such a way to
account the number of molecules in a new phase.
It is absolutely sufficient to take the following
expression
$$
\int_{x_{k-i}}^{x_{k-i+1}} f_i \rho^3 d \rho =
N_i  x_{k-i}^3
$$
which is valid at  $k-i \gg 1$.
In a whole quantity of substance it is sufficient
to take into account only droplets with $k-i \gg 1$.
The relative weight of dismissed terms will be
small.

Then for the total number of molecules in droplets at
interval number $k$ we have the following
expression
$$
Q_k = \sum_{i=1}^k N_i x_{k-i}^3
$$
where $x_{k-i}$ is appropriate coordinate or
$$
Q_k \sim \sum_{i=1}^k N_i  \hat{\Delta}^3 (k-i)^3
$$

This representation is important because now the
note  in
\cite{Koll}
after eq. (15)
isn't necessary. This note stated that the
probability for
$N_i$
to deviate from  $\bar{N}$
is very low. This note
is doubtful because namely these deviations are
the base for stochastic effects. Now there is no
need in this note.

The next step is to build \cite{Koll}
two cycle construction for nucleation period.
During  the first cycle
the main consumers of vapor appeared and during
the second cycle  they rule a process of
formation of all other droplets. In
\cite{Koll}
it is supposed that during the first cycle a
vapor depletion is negligible and during the
second cycle new droplets are absolutely governed
by droplets from the first cycle.  Now we shall
analyze an effectiveness of such procedure.

We analyze an equation
$$
\psi = \int_0^z (z-x)^3  \exp(-\psi)) dx
$$
The first iteration \cite{PhysRevE94} is
practically precise solution and it gives the
number of droplets
$$
N_{tot} =
\frac{1}{4} \frac{4^{1/4} \pi \sqrt{2}}{\Gamma
(3/4)} = 1.2818
$$

A model solution requires that until  $z=p$
there will be no depletion of vapor and
then only droplets formed before $z=p$
will consume vapor. Then for a total number of
droplets we have an expression
$$
N_{tot \ appr} = p + \int_p^{\infty}
\exp(-\frac{1}{4} x^4  + \frac{1}{4} (x-p)^4 ) dx
$$
A ratio $N_{tot \ appr}/ N_{tot}$
is given at fig. 4

\begin{figure}[hgh]

\includegraphics[angle=270,totalheight=10cm]{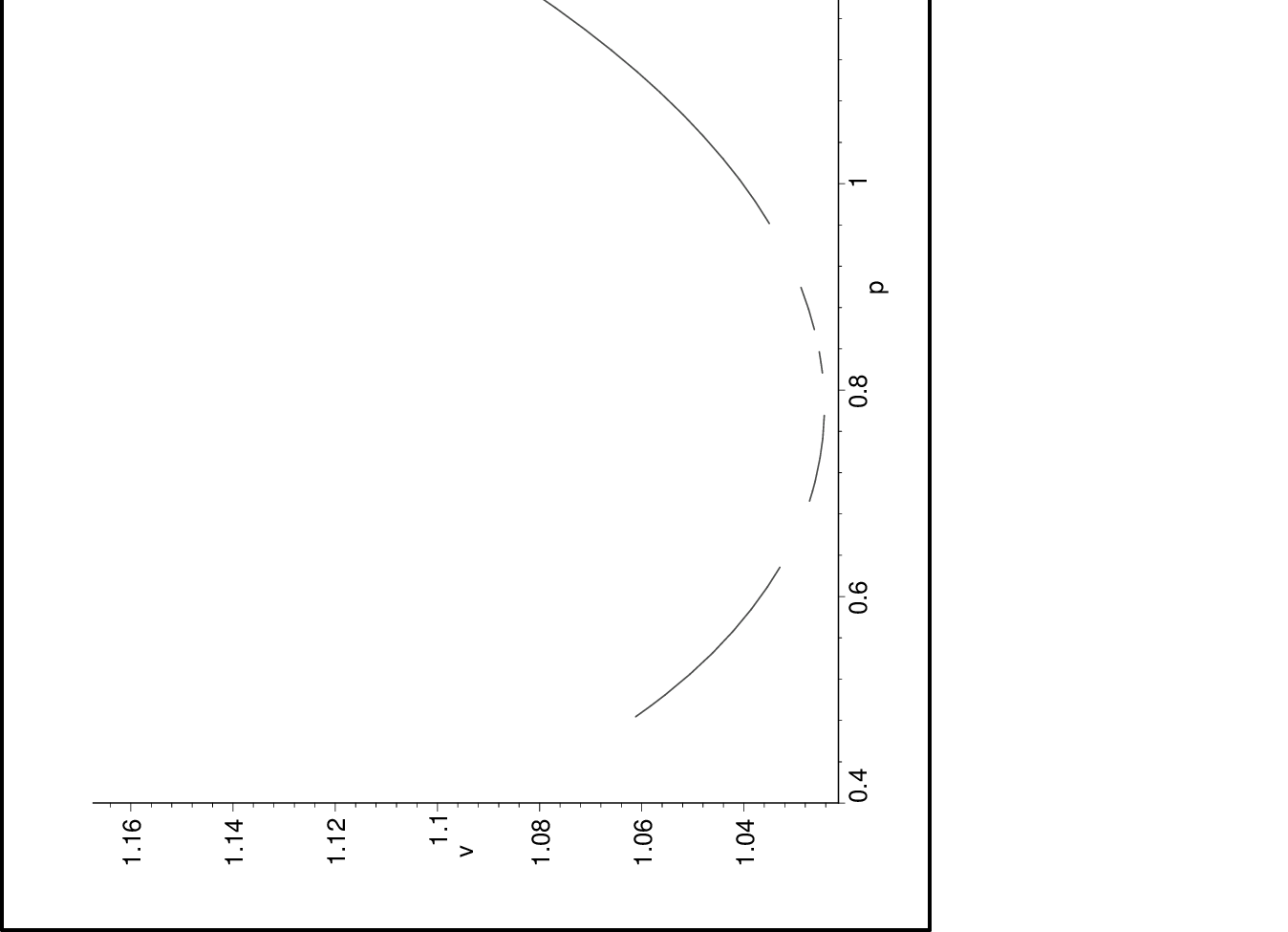}

\begin{caption}
{
The ratio of mean numbers of droplets }
\end{caption}

\end{figure}

Now it is clear that in \cite{Koll}
the value of parameter of separation into two
cycles was not chosen in a good style (at least
from the point of view in the theory with
averaged numbers).
We can also stress the smooth dependence
$N_{tot \ appr}$ on $p$.

We shall study the probability $P_k$
of formation  of stochastic number  $N_k$
of droplets at the first  $k$
 elementary intervals \cite{Koll}.

Our constructions now resemble  \cite{Koll}
but there is one
essential difference. When we get a formula analogous
to  (30) \cite{Koll} we have no necessity to
linearize expression with respect to
$(N_i - \bar{N_i} )/ \bar{ N_i}$,
where
$N_i$
is a stochastic number of droplets formed at interval
$i$,
$\bar{N_i}$ is a mean number of droplets formed
at interval $i$ (it is a function of stochastic
numbers of droplets at preceding intervals).
This linearization can not take place because
a  ratio $(N_i - \bar{N_i}) / \bar{ N_i}$
can be zero or can attain huge value (with low
probability).
It is more simple and more justified to linearize
expression on
$\sum_i \rho_i^3 (N_i - \bar{N_i}) / \bar{ N_i}   $
where $\rho_i$
is a linear size of droplets formed at interval $i$
(all of them have approximately the same size).
Really, due to summation variations of
$\sum_i \rho_i^3 (N_i - \bar{N_i}) / \bar{ N_i}   $
are much smaller than variations of
 $(N_i - \bar{N_i}) / \bar{ N_i}$.

Variations of $(N_i - \bar{N_i}) / \bar{ N_i}$
would be small only at very big numbers of droplets
$N_{tot}$.
 One can get
$$ (N_i - \bar{N_i}) / \bar{ N_i} \sim \bar{N_i}^{1/2} $$
$$ N_{tot} \sim M \bar{N_i}$$
$M$ is a number of elementary intervals.
So, the  theory with linearization  proposed in
\cite{Koll}
would be well justified only in a region where
the result can be obtained on the base of
averaged characteristics. But the linearization
proposed here leads to the same numerical
expressions. For dispersion of the total
distribution one can get
$$
D^{\infty} = \tilde{N^{\infty}}
(1-\frac{\beta}{\alpha})
$$
where
$$
\beta = \beta_1 - \beta_2
$$
$$
\beta_1 =
8
\int_{1/2}^{\infty} d \xi \int_0^{1/2} d \tau (\xi -
\tau)^3
\exp(-\xi^4)
$$
$$
\beta_2 =
16
\int_{1/2}^{\infty} d \xi
\int_{1/2}^{\infty} d \eta
\int_0^{1/2} d \tau
(\tau- \xi)^3
(\tau-\eta)^3
 \exp(-\xi^4)
 \exp(-\eta^4)
$$
$$
\alpha = \int_0^{\infty} dx \exp(-x^4)
$$

In two cycles construction the value of $\alpha$,
which is proportional to the total number of
droplets has to be reconsidered and recalculated
on the base of two cycles. Then we have to use
instead of previous
$\alpha
\equiv \alpha_0$
a new value
$$
\alpha = \alpha_1 \equiv
1/2 + \int_{1/2}^{\infty}
\exp(-x^4 + (x-1/2)^4)
$$

We shall use parameter $p$
of separation of two cycles and we shall
calculate
$\alpha_1$ as
$$
\alpha_1 \equiv
p + \int_{p}^{\infty}
\exp(-x^4 + (x-p)^4)
$$
Then according to fig. 4 we see that
 the ratio    $\alpha_0 / \alpha_1$ is greater
 than $1$ and
  $\alpha_1$ is greater than $\alpha_0$.
Here we see that
two-cycles construction is approximate one.
Result for  $D^{\infty}$
will differ from
\cite{Koll}
and will be
$$
D^{\infty}_e = \tilde{N^{\infty}}
0.69
$$
instead of
$$
D^{\infty}_f = \tilde{N^{\infty}}
0.67
$$
as it is stated in \cite{Koll}.
The value $D_e^{\infty}$ differ from a real
result more than one tenth. So, the new theory is
necessary.

Now it is necessary to decide what $p$ shall we
choose. At arbitrary $p$ the expression for $\beta$
will be the same but for $\beta_1$  è $\beta_2$
we have
$$
\beta_1 =
8
\int_{p}^{\infty} d \xi \int_0^{p} d \tau (\xi -
\tau)^3
\exp(-\xi^4)
$$
$$
\beta_2 =
16
\int_{p}^{\infty} d \xi
\int_{p}^{\infty} d \eta
\int_0^{p} d \tau
(\tau- \xi)^3
(\tau-\eta)^3
 \exp(-\xi^4)
 \exp(-\eta^4)
$$
We have to reconsider expression for $\alpha$.

After calculations we have for dispersion as function of
$k \equiv p$ the following fig. 5

\begin{figure}[hgh]

\includegraphics[angle=270,totalheight=10cm]{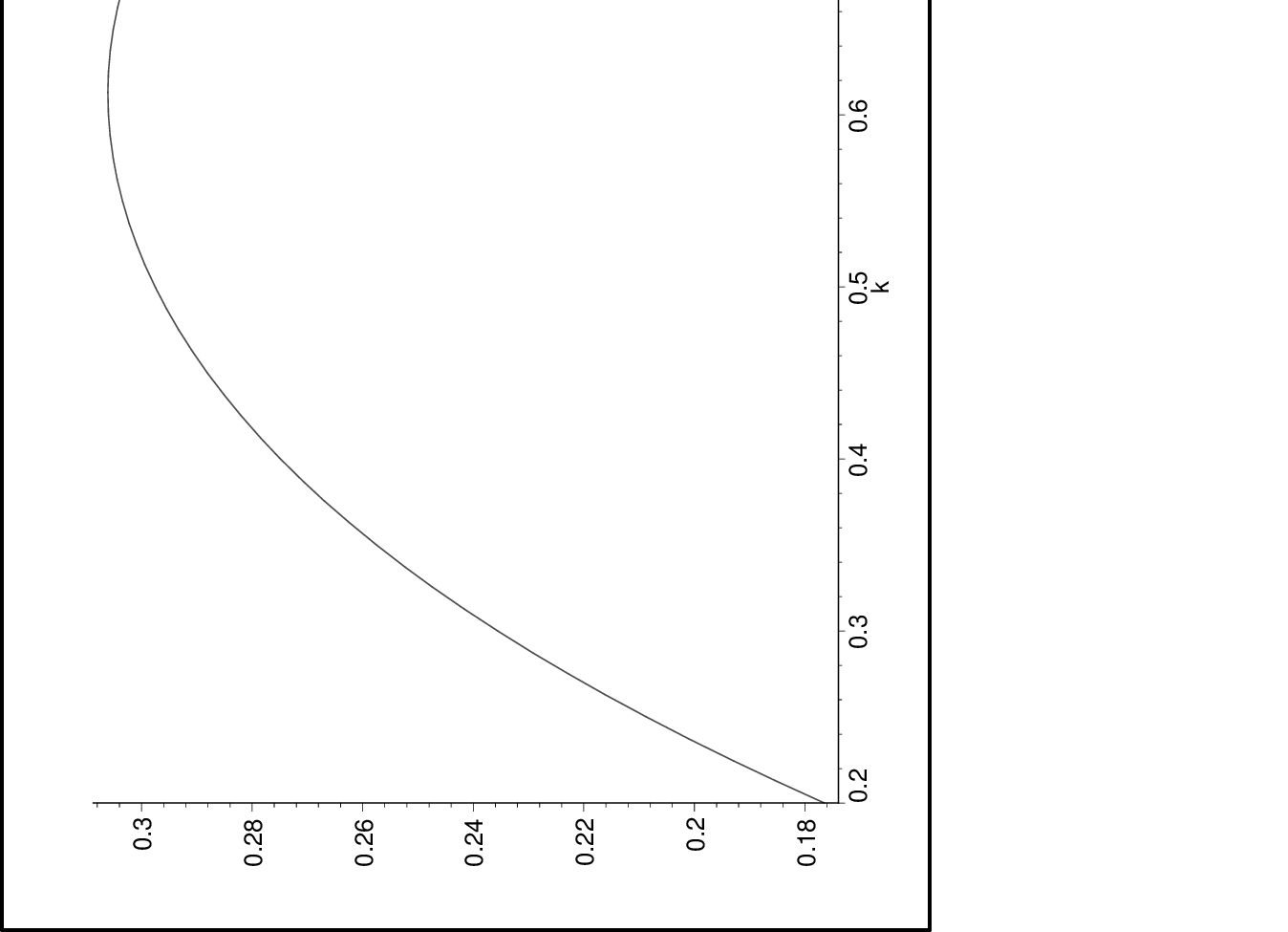}

\begin{caption}
{Dispersion as function of  $k$}
\end{caption}

\end{figure}

A minimal dispersion will be at $p=0.6$.
This is the true value of $p$ because this value
corresponds to the sense of minimal work.
Dispersion will be
$$
D^{\infty} = 2 \tilde{N^{\infty}}
0.66229
$$

We shall make our result more precise.  We see
that
$\beta_1$ and $\beta_2$
are the first two terms of some series.
We don't know other terms, but
it is reasonable to suppose that this series
resembles geometric progression with denominator
$\beta_2 / \beta_1$.
This leads to dispersion
$$
D^{\infty} = 2 \tilde{N^{\infty}}
0.64107
$$
As it follows from fig.6 the   value of extremum
 remains
$p=0.6$


\begin{figure}[hgh]

\includegraphics[angle=270,totalheight=10cm]{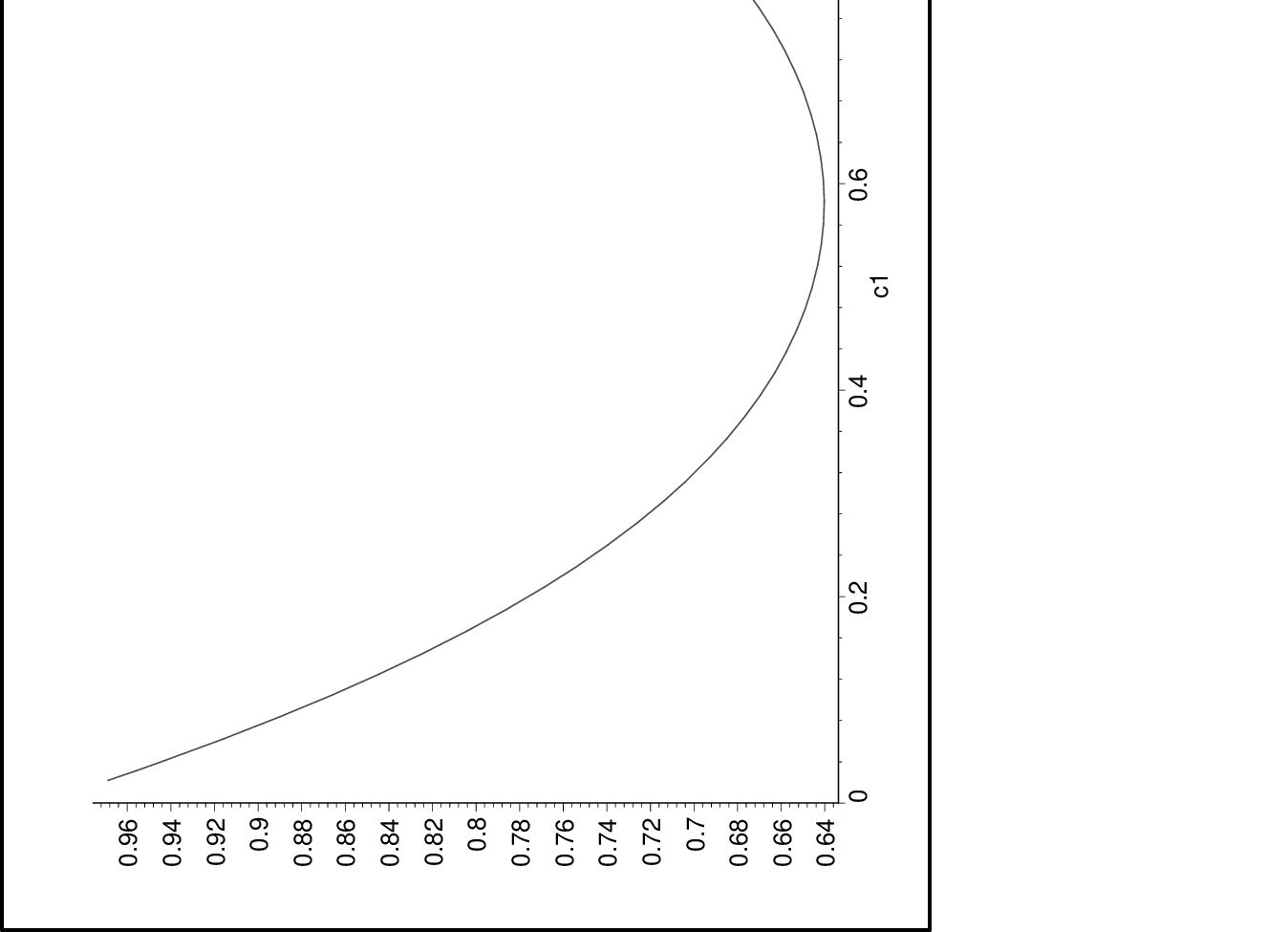}

\begin{caption}
{Dispersion as function of $c1 \equiv p$}
\end{caption}

\end{figure}

An absence of extremum shift is important and is
necessary for this approach to be a self consistent.

Now we shall make the value of dispersion more
accurate
Due to the similarity of nucleation the first
cycle doesn't differ from the whole period.
   Function
   $\beta$
for the first cycle will be
$$
\beta = \beta_1 - \beta_2
$$
$$
\beta_1 =
8
\int_{p_1}^{p} d \xi \int_0^{p_1} d \tau (\xi -
\tau)^3
\exp(-\xi^4)
$$
$$
\beta_2 =
16
\int_{p_1}^{p} d \xi
\int_{p_1}^{p} d \eta
\int_0^{p_1} d \tau
(\tau- \xi)^3
(\tau-\eta)^3
 \exp(-\xi^4)
 \exp(-\eta^4)
$$
Calculations for $p_1 = 0.6*0.6 = 0.36$ and  $p=0.6$
give $\beta = 0.0255$.

Now we shall reconstruct $\beta_{eff}$ based on
$
D^{\infty}_2 = 2 \tilde{N^{\infty}}
0.64107
$
We have $\beta_{eff} = (1 -
D^{\infty}/\tilde{N^{\infty}})/\alpha$  where
$\alpha = 0.9092$. Calculations give $\beta_{eff} =
0.3268$.
One has to add
$\beta = 0.0255$
which leads to  $\beta = 0.35156$.
For this value of $\beta$
the value of dispersion will be
$$
D^{\infty}_3 = 2\tilde{N^{\infty}}
0.61332
$$
This is our final result.

When we consider the dispersion  which differs from
$
D^{\infty}_0 = 2\tilde{N^{\infty}}
$
we have the shift of result
$$
\delta D = D^{\infty}_3- D^{\infty}_2
$$
This shift has to lead to a variation of
dispersion in a fist cycle and to a variation of
the total dispersion.
Having linearized this effect we get to the final
value of dispersion
$$
D_4^{\infty} = D_2^{\infty} +
\delta D \frac{D_2^{\infty}}{D_3^{\infty}}
$$
and after calculations
$$
D^{\infty}_3 = 2\tilde{N^{\infty}}
0.62309
$$
This value practically coincides with the
previous value.

Another way to make results more precise is
take into account the shift of dispersion
directly in initial formulas. Having written for
the dispersion correction in the first cycle
$$
D^{\infty}_3 = 2\tilde{N^{\infty}}
\gamma
$$
with parameter $\gamma$,
we get
for the final distribution
$$
P^{(k)} (M^{(k)} \sim
\int_{\infty}^{\infty}
dN_1 dN_2 ... d N_P
\prod_{i=1}^{P} \exp(-\frac{(N_i  - \bar{N_1})^2}{2
\gamma \bar{N_1}})
$$
$$
\exp[\frac{[N^{(k)} - \tilde{N^{(k)}} -
\sum_{j=1}^{P} a_j (N_j - \bar{N_j})]^2}
{2(\tilde{N^{(k)}}-P \bar{N_1})}]
$$
where
$P$
is the number of elementary intervals until argument
$p$,
$\tilde{N^{(k)}}$
is the number of droplets calculated on the base
of the theory with averaged characteristics,
$\bar{N_i}$
the mean number of droplets formed during interval number
$i$
with account of fluctuations from previous intervals.
The values $a_i$
 are given by
$$
a_i = 1 - \sum_{j=P+1}^{k}
\frac{\exp(-j^4/M^4)}{M^4} 4 (j-i)^3
$$
and $M$
is the total number of intervals.

Having fulfilled integration
$\int_{\infty}^{\infty}
dN_1 dN_2 ... d N_P$,
we get for a limit value of dispersion
$$
D^{\infty} =  2 \tilde{N^{\infty}} ( 1-
\frac{p(1-\gamma)}{\alpha} - \frac{\gamma
\beta}{\alpha})
$$
 which leads to equation on $\gamma$,
which can be easily solved
 $$
\gamma =
\frac{1 - \frac{p}{\alpha}}
{1+ \frac{\beta}{\alpha}- \frac{p}{\alpha}}
$$
Calculations lead to
$$
\gamma (p=0.6) = 0.5083
$$

This result is very strange. It radically differs
from the previous one. The reason of an error is
that the duration of the first cycle is limited
by
$p$.
So, we have to limit the duration of a whole
period. The limit is, evidently,
 $\sim 1 $.
So, we have to recalculate $\beta$
as
$$
\beta = \beta_1 - \beta_2
$$
$$
\beta_1 =
8
\int_{p}^{1} d \xi \int_0^{p} d \tau (\xi -
\tau)^3
\exp(-\xi^4)
$$
$$
\beta_2 =
16
\int_{p}^{1} d \xi
\int_{p}^{1} d \eta
\int_0^{p} d \tau
(\tau- \xi)^3
(\tau-\eta)^3
 \exp(-\xi^4)
 \exp(-\eta^4)
$$
Now we shall show a dependence  $\beta_{initial}$
on
 $p$ at fig. 7.


\begin{figure}[hgh]

\includegraphics[angle=270,totalheight=10cm]{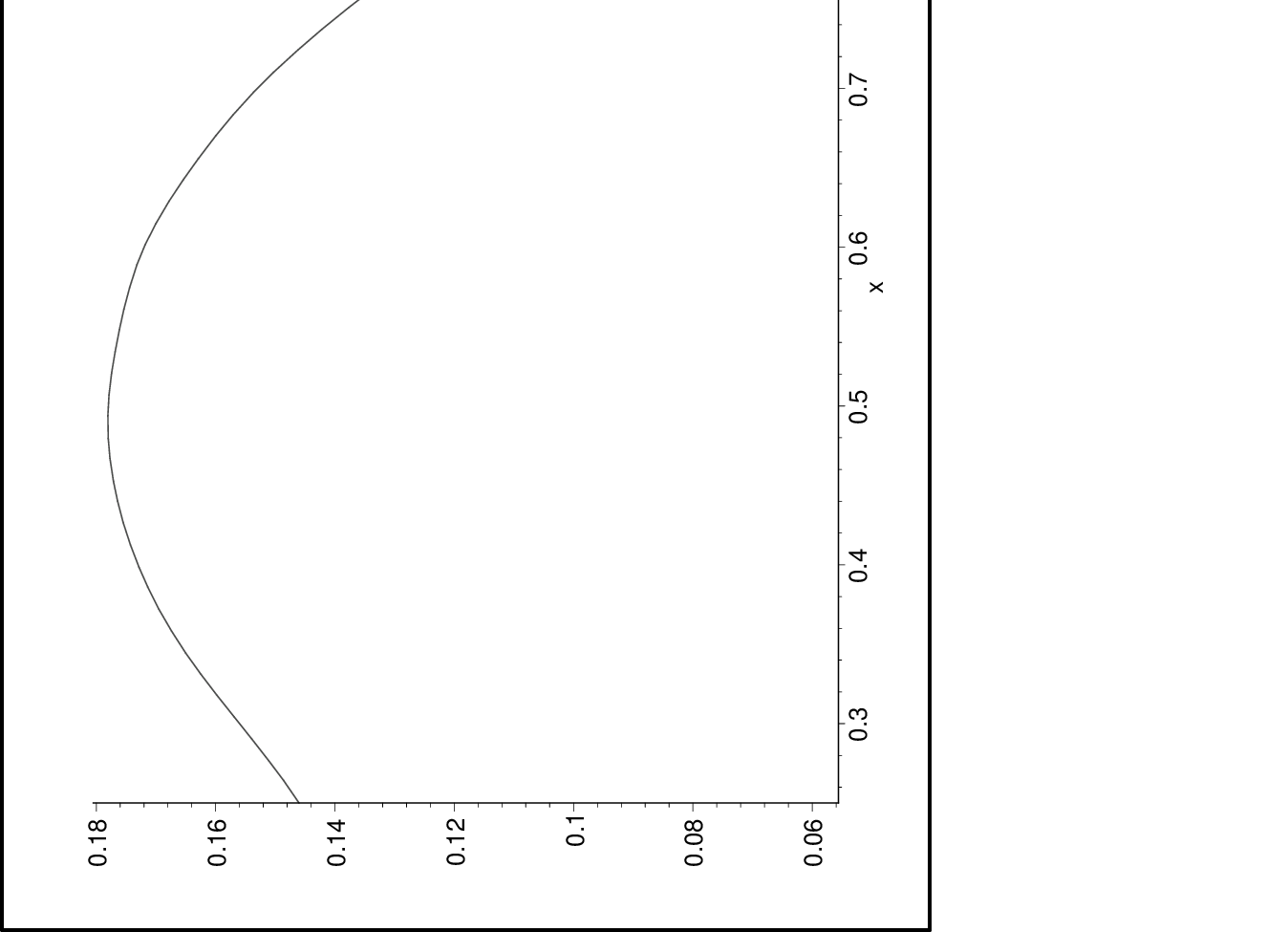}

\begin{caption}
{ $\beta_{initial}$ as a function of $ p$}
\end{caption}

\end{figure}

Calculations give $\beta_{initial} = 0.17$
and for the final dispersion
$$
D^{\infty}_3 = 2\tilde{N^{\infty}}
0.6436
$$

This value coincides with a previous approach. It
can be made more accurate by a geometric
progression summation spoken above.

\section{Numerical results}

Numerical simulation of nucleation can be done by
the following method. We split the nucleation
interval into many parts (steps). At every step a
droplet will be formed or not. The probability to
appear must be rather low, then we have the
smallness of probability to have two droplets at
the same interval.  This means that the interval
is "elementary".

The process of formation is simulated by a
random generator in a range $[ 0, 1]$. If a
generated number is smaller than a threshold
parameter  $u$,
then there will be no formation of a droplet. If it is
greater than a threshold, we shall form a droplet.
As a result we have spectrum $\hat{f} $
of droplets sizes. Now it is a chain of $0$ and $1$.
The parameter $u$
descends according to macroscopic law
$$
u = u_0 \exp(- \Gamma G /\Phi )
$$
from a theory with averaged characteristics (it
is based only on a conservation law without any averaging
 and can be used).
Here
$$
\Gamma = \frac{d F_c}{d \Phi} \sim \nu_c
$$
$
\Phi
$
is the initial supersaturation,
$F_c$
is a free energy of critical embryos formation,
 $\nu_c$
is a number of molecules inside a critical
embryo,
$G$
is the number of molecules in a new phase taken in
units of
a molecules number density in a saturated vapor.
By renormalization one can take away all parameters
except $G$.

To simplify calculations radically
one can use the following representation
\cite{PhysRevE94} for
$G$:
$$
G = z^3 G_0 - 3  z^2 G_1 + 3 z G_2 - G_3
$$
where   $z$
is a coordinate of a front of spectrum,
and $G_i$
are given by
$$
G_i = \int_0^z \hat{f}(x) x^i dx
$$
We needn't to recalculate $G_i$,
but can only ascend the region of integration,
having added to integrals
$z^i\hat{f}(z)  dx$ at every step.

Our results are given below. The interval is split into
$30000$
parts. Parameter
 $u_0$
have been varied from  $0$ up to $1$
which leads to a different
number of droplets. It is clear that the limit
values are not good: at  $0$
there are no droplets in the system, at $1$
our intervals are not elementary.
At every $u_0$
results were averaged over $1000$ attempts.

Shifts of droplets numbers are
drawn at fig.8  as a function of
$\ln \tilde{N^{(\infty)}}$

\begin{figure}[hgh]

\includegraphics[angle=270,totalheight=10cm]{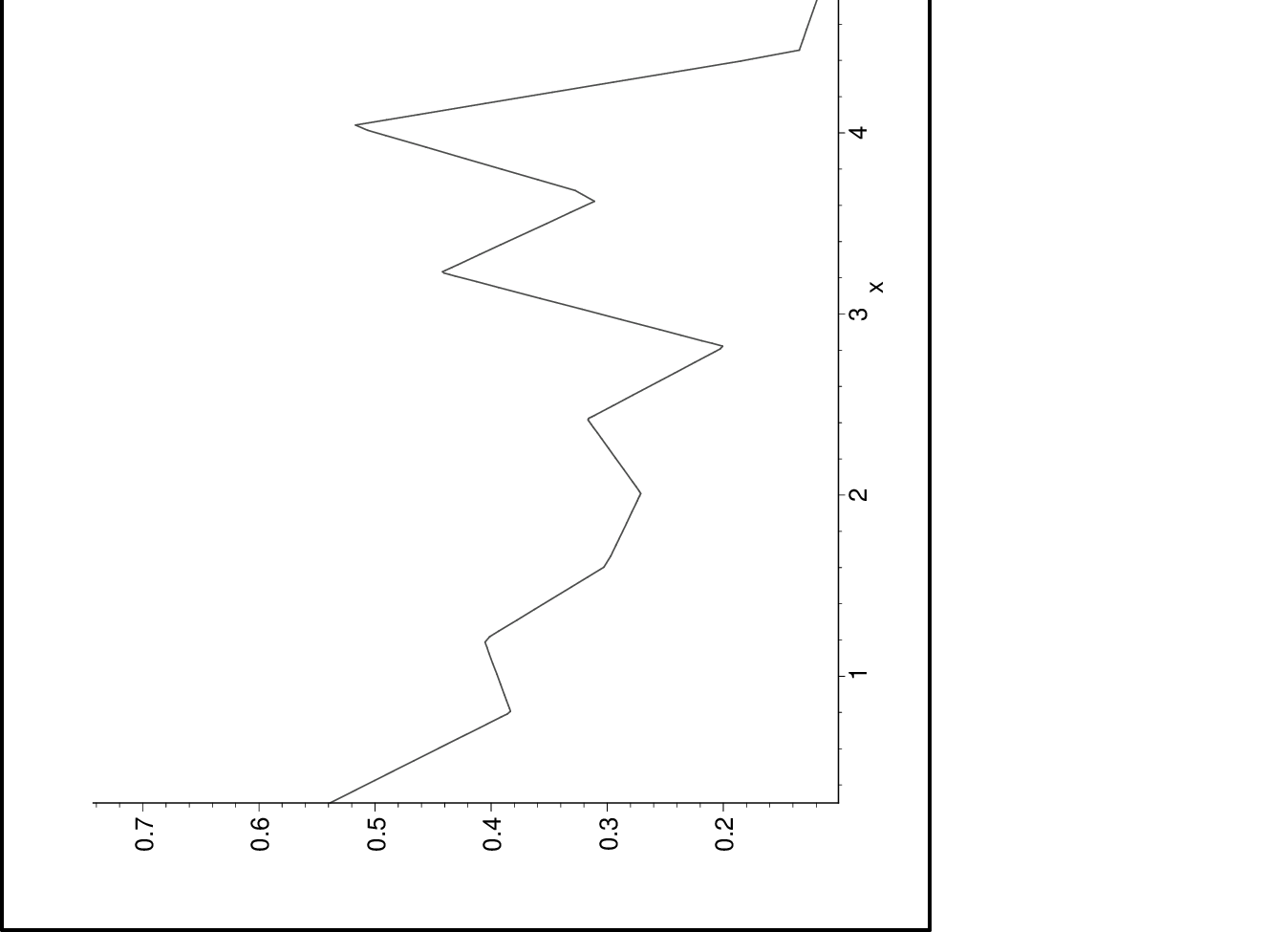}

\begin{caption}
{
Shifts of droplets numbers as a function of
 $\ln \tilde{N^{(\infty)}}$}
\end{caption}

\end{figure}
It is seen that an analytical result about
negligible value of corrections is correct.

Dispersion as a function of $\ln
\tilde{N^{(\infty)}}$
is shown at fig. 9.


\begin{figure}[hgh]

\includegraphics[angle=270,totalheight=10cm]{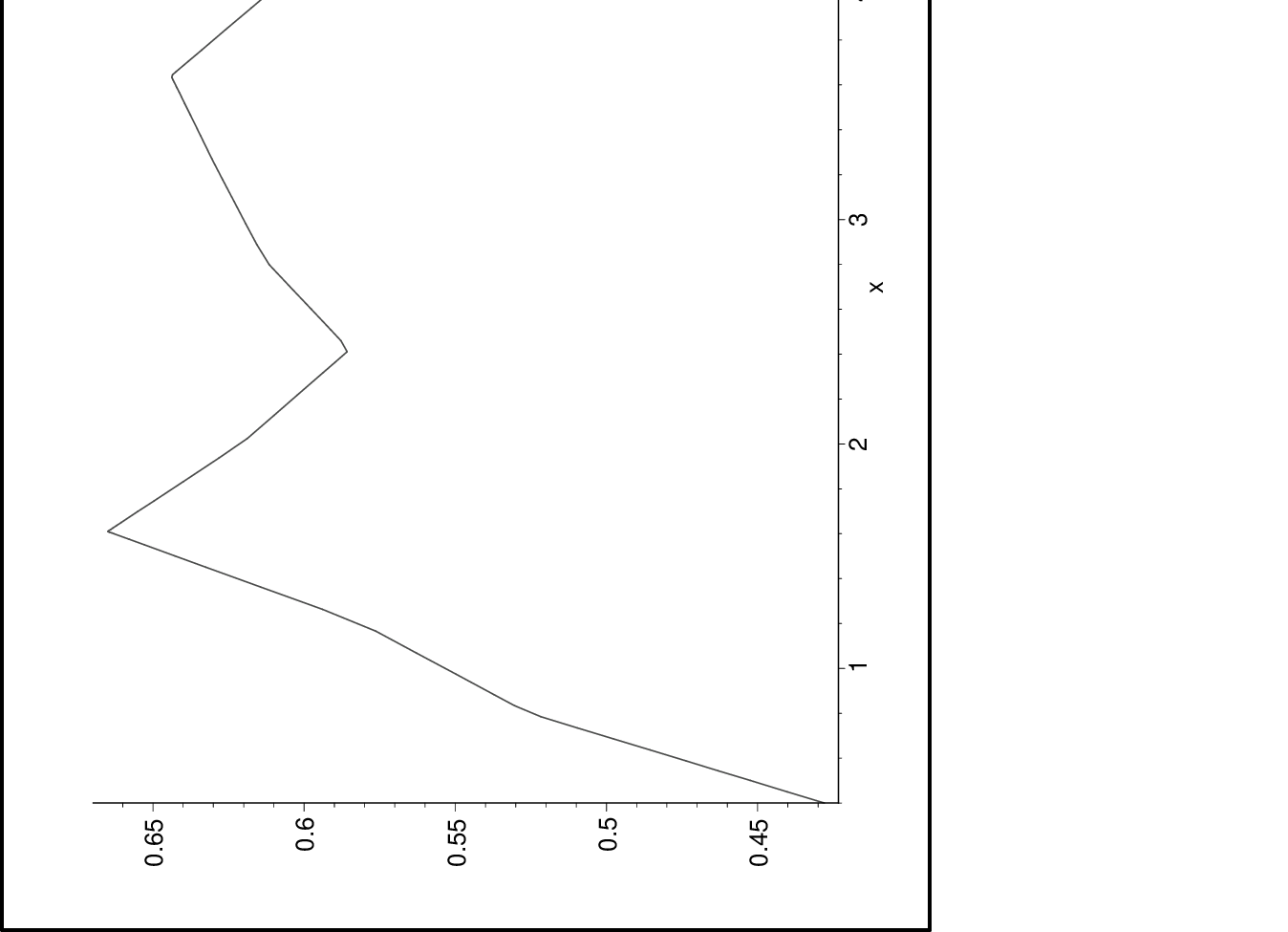}

\begin{caption}
{  $\gamma$
as a function of  $\ln \tilde{N^{(\infty)}}$}
\end{caption}

\end{figure}

It is seen that analytical value of dispersion
coincides with numerical simulation.
The ends of the curve correspond to a zero
number of droplets and to a giant number of
droplets when the elementary intervals are not
elementary  and have to be thrown out.

Stochastic effects in dynamic conditions
\cite{PhysRevE94} can be
analyzed by the same method. We needn't to
describe it here. Numerical results are drawn
below. Fig. 10 shows the shift in the number of
droplets. It is small.
  Dispersion is drawn at fig. 11 (i.e. the value
  of
$\gamma$).
It is greater than in the case of decay.

\begin{figure}[hgh]

\includegraphics[angle=270,totalheight=10cm]{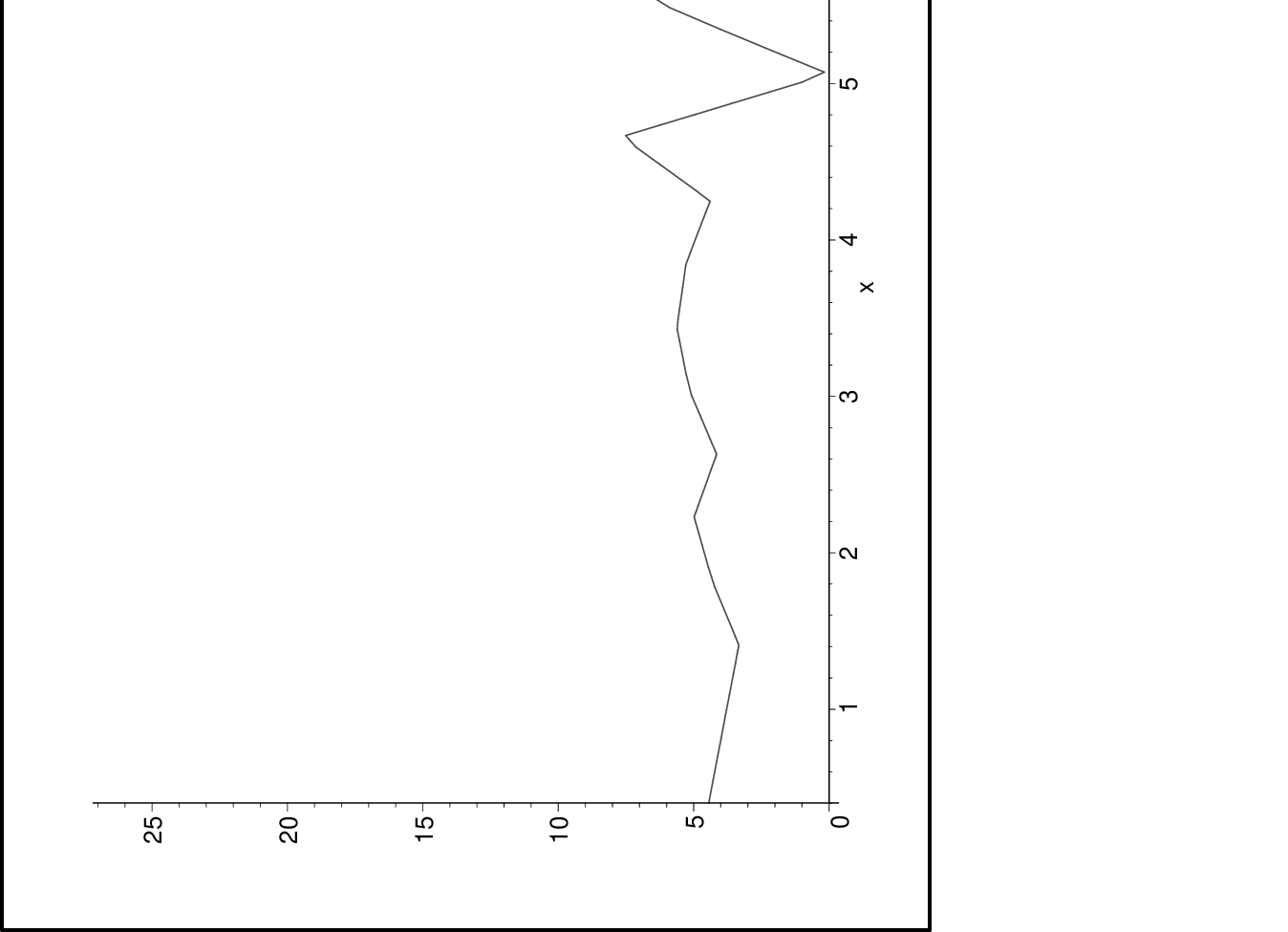}

\begin{caption}
{
Shift in a droplets number as a function of
 $\ln \tilde{N^{(\infty)}}$ for dynamic conditions}
\end{caption}

\end{figure}


\begin{figure}[hgh]

\includegraphics[angle=270,totalheight=10cm]{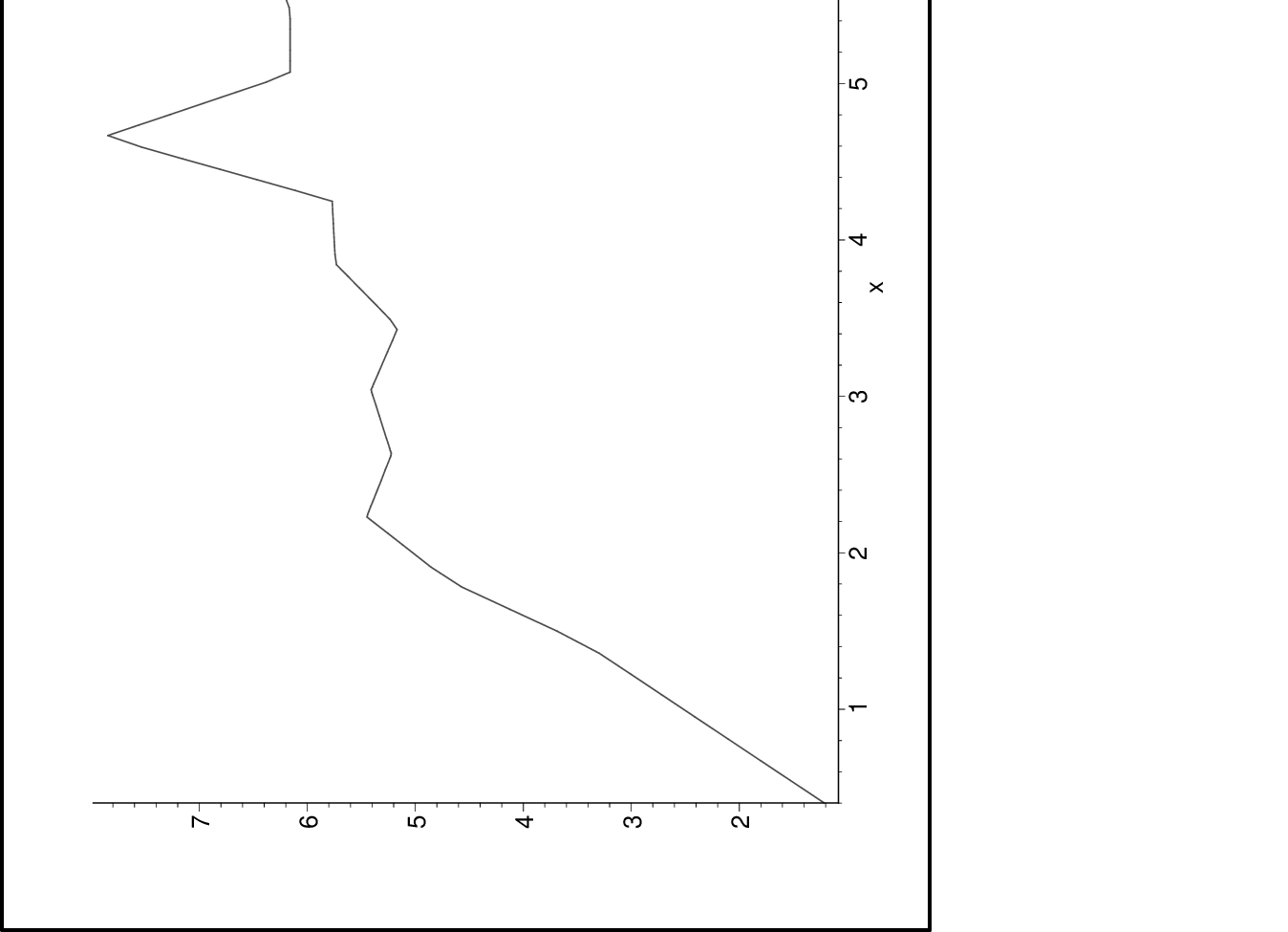}

\begin{caption}
{ $\gamma$
as a function of  $\ln \tilde{N^{(\infty)}}$
for dynamic conditions}
\end{caption}

\end{figure}

The physical reasons for the smallness of the
droplet number shift for decay and for dynamic
conditions will be different.

For decay the reason is the following. The system
wait the first droplet as long as necessary.
Actually the time for kinetics of
this system is $t(G)$ with
no connection with real time (certainly, the rate
of nucleation has such connection).
This phenomena is the reason for a smallness.

In dynamic conditions there is a time dependent
parameter - the intensity of external source. So,
 there is no such a reason.

But here in the theory with averaged
characteristics there is a property of a weak
dependence on microscopic corrections for a free
energy \cite{PhysRevE94}. The same is valid also
for fluctuation deviations. So there will be a
weak effect of stochastic nucleation.

Because the reasons for smallness of effect in
decay and dynamic conditions are different it is
interesting to see whether they continue to act
when
supersaturation is stabilized at some moment.
Analytical results shows that the will be an
overlapping of two reasons.

Really, if stabilization
takes place at the period where the main
consumers of vapor are going to appear then the
majority of droplets appear in the situation when
there is no influence on the system. Then the
situation for these droplets resembles decay
conditions (and may be even better because the
external supersaturation \cite{book1} is going to
decrease). So the reason for the decay situation
works here.

If stabilization takes place at the
second cycle, then the behavior of
supersaturation is governed by droplets formed in
dynamic conditions and we have here the reason
for smallness in dynamic conditions. In both
situations the effect is small.
Numerical results
confirm this conclusion.

The main result of this publication is a correct
definition of all main characteristics of
stochastic nucleation.  It is shown that the main
role in stochastic effects belongs to all
droplets, but not to the main consumers of vapor.
Only the property of the nucleation conditions
similarity allows us to solve the problem of
account of all influences during the nucleation
period.

When all disadvantages of \cite{Koll}, \cite{Vest}
are shown it is clear that these theories can not
be considered as a solid base for nucleation
investigation.

But why results obtained in  \cite{Koll},
\cite{Vest}
are so close numerically to real values.
The reason is that on a level of averaged
characteristics there is a universality of
nucleation process. So, the errors of
\cite{Koll}, \cite{Vest}
cannot lead to a qualitatively wrong results.

One has to stress that all corrections obtained
in this paper are also universal ones. Certainly,
they are some coefficients in decompositions and
the functional form is prescribed now (contrary
to the theory with averaged characteristics).

It seems that all effects considered here are
negligible. For simple systems it is really true.
But for systems with more complex
kinetic behavior these effects can be giant. One
of such systems is already described
theoretically and this description
will be presented soon in
a separate publication.

In diffusion regime of
droplets growth one has to use another
approach based on
\cite{PhysicaA},
\cite{PhysRevE2001}. In  \cite{PhysRevE2001}
an explicit description of nucleation with
account of stochastic effects was constructed.

A nucleation with growing volumes of interaction
will be presented in the next paper.

\end{document}